\documentstyle[12pt,openbib]{article}
\def\be{\begin{equation}}
\def\ee{\end{equation}}

\begin{document}

\title{\bf Members of the double pulsar system PSR J0737-3039 : neutron stars or strange stars ? }

\author{Manjari Bagchi $^{1}$, Jishnu Dey $^{2,\ddagger, *}$, Sushan Konar$^{3}$ \\ Gour Bhattacharya $^{2}$, Mira Dey $^{2,\ddagger,*}$ }

\maketitle

\begin{abstract}

One interesting method of constraining the dense matter Equations of State is to measure the advancement of the periastron of the orbit of a binary radio pulsar (when it belongs to a double neutron star system). There is a great deal of interest on applicability of this procedure to the double pulsar system PSR J0737-3039 (A/B). Although the above method can be applied to PSR A in future within some limitations, for PSR B this method can not be applied. On the other hand, the study of genesis of PSR B might be useful in this connection and its low mass might be an indication that it could be a strange star.

\end{abstract}

\vskip .5cm



\vskip .5cm

\noindent {$^1$ Tata Institute of Fundamental Research, Colaba, Mumbai 400005, India \\
$^2$ Dept. of Physics, Presidency College, Kolkata 700073, India \\
$^3$  Dept. of Physics, Indian Institute of Technology, Kharagpur, India\\
$^\ddagger$ Associate, IUCAA, Pune, India \\
$^*$ Supported by DST Ramanna grant, Govt. of India.\\
}

\newpage

\section{Introduction}

It was argued by Witten \cite{witten} that {\em strange quark matter},
composed of $u$, $d$ and $s$ quarks, is more bound than nuclei at zero 
pressure and is therefore the true ground state of matter. Ever  
since the advent of this hypothesis the existence and the observational  
characteristics of the {\em  strange  quark stars} \cite{alck86, haen86} 
are being debated.

Unfortunately, strange stars and neutron stars have a number  of
almost similar physical properties like the range of masses, stable rotation
periods, cooling history and so on, but strange stars are more compact than 
neutron stars. Because of this, quite a few compact objects were speculated 
to be strange stars but there has been no conclusive evidence for  the case 
or against it. For example, RX J1856.5-3754 has been the subject of such a 
recent debate because of certain peculiarities of its  thermal X-ray 
properties \cite{drake02, hend07}.

Understandably the most convincing proof for a strange star would
come from a direct indication of the equation of state and we believe 
that the double neutron systems (DNS) might provide us with that
answer. Precise measurements of a number of parameters of DNS systems
allow for the study of a variety of phenomena like the effects of 
general relativity, the emission of gravitational waves or the
intricacies of the binary evolution. In particular, in the case  of PSR
J0737$-$3039, the only double pulsar system \cite{burgay, lyne} detected so far, it is possible to measure many more properties like the mass ratio of the two neutron stars directly obtainable from pulsar timing. The two pulsars are named as J0737$-$3039A and J0737$-$3039B. The values of spin period and magnetic field of these two are $P~=~22.7$ ms, $B~=~6.4 \times 10^9$ G for pulsar A and  $P~=~2.77$ s  and $B~=~1.6 \times 10^{12}$ G for pulsar B.  Apparently A is a typical recycled pulsar and B is the one born second.   The periods $P_{A,B}$ and their time derivatives provide upper limits for the lifetimes  $t_{A} \approx 210$ Myr and $t_{B} \approx 50$ Myr. The estimated separation (sum of the semi major axes) was $R \sim 10^{12}$ cm and the eccentricity was $e~\sim~0.11$ when pulsar B was born. The present values of separation and eccentricity are $8.8~\times 10^{11}$ cm and 0.0877775(9). As the values of R and e does not change considerably with time, the analysis of the origin of PSR J0737$-$3039B is insensitive to the exact age of the pulsar.

 Formation models of PSR J0737$-$3039 were considered by various authors \cite{dewi04, pod, ps05}.  Dewi \& van den Heuvel \cite{dewi04} proposed that J0737$-$3039 has originated from a close helium star plus neutron star (HeS–NS) binary in which at the onset of the evolution the helium star had a mass in the range 4.0 $-$ 6.5 $M_{\odot}$ and an orbital period in the range 0.1 $-$ 0.2 day. In their model,  a kick velocity in the range 70 $-$ 230 $km~/ sec$ must have been imparted to the second neutron star at its birth. Podsiadlowski $et~al.$  \cite{pod} proposed an electron capture supernova scenario. Using the observational indications that the binary system will end up within 50 pc from the galactic plane with $0.088<e>0.14$ and with a transverse velocity less than 30 km/s, provided 50 Myrs in the past the progenitor system had a circular orbit, Piran \& Shaviv \cite{ps05} estimated that it is most likely that the progenitor of PSR J0737$-$3039B had a mass of $1.55 \pm 0.2~M_\odot$ with a low kick velocity 30 km/s. The progenitor may well have been a NeOMg star.  Willems $et~al.$ \cite{willems} used population synthesis method to estimate the most likely radial velocity of the binary. They favored the standard formation scenario with reasonably high kick velocities (70-180 km/s) and pre-SN mass $m_{2,i}$ of at least $2~M_{\odot}$ which fits with a Helium star progenitor, but acknowledged that for very small transverse velocities ($\sim$ 10 km/s) lower mass progenitors are allowed or even favored. 

\section{The nature of the members of PSR J0737-3039}

The system is rich in observational phenomena, including a short radio eclipse of A by B and orbital modulation of the radio flux of B affected by A. Much radio pulse data from both pulsars has already been accumulated. High energy observations could also be important to these studies since both stars also emit X-rays and they are now being detected recently. 

\subsection{radio observations}

The best answer to our question regarding the equation of state would come from the measurement of the moment of inertia of individual stars in this system. The moment of inertia can be determined by measuring the extra advancement of the periastron of the orbit above the standard post-Newtonian one due to the spin-orbit coupling \cite{ds} using timing analysis of radio data. Periastron advance (which is the observable quantity) for a DNS system is given by :
\begin{equation}
\dot{\omega}~=~\frac{3 \beta_0^2~ n}{1-e^2}\left[1+f_0 \beta_0^2 -(g_{s1} \beta_{s1}\beta_0+g_{s2} \beta_{s2}\beta_0) \right]
\label{eq:per_adv}
\end{equation}
where the first term inside the square bracket represents 1PN term, the second term term inside the third bracket represents 2PN term, and the third term inside the third bracket represents spin-orbit coupling (SO). The parameters used in the above expression are defined as :
\begin{equation}
\beta_0~=~\frac{(GMn)^{1/3}}{c}
\label{eq:beta0}
\end{equation}

\begin{equation}
\beta_{sa}~=~\frac{c I_a \omega_a}{G m_a^2}
\label{eq:betas}
\end{equation}

\begin{equation}
f_0~=~\frac{1}{1-e^2}\left( \frac{39}{4}x_1^2+\frac{27}{4}x_2^2+15 x_1 x_2 \right) -\left( \frac{13}{4}x_1^2+\frac{1}{4}x_2^2+\frac{13}{3} x_1 x_2 \right)
\label{eq:f0}
\end{equation}

\begin{eqnarray}
g_{sa}~=\nonumber~\frac{x_a \left(4 x_a+ 3x_{a+1}\right)}{6(1-e^2)^{1/2}sin^2 i}\left[ (3 ~sin^2 i-1)~{\it \bf   k~.~s_a}+cos~ i~ {\it \bf K_0~.~s_a} \right]   \\ 
\label{eq:gs}
\end{eqnarray}

where $G$ is the gravitational constant, $I_a$ is the moment of inertia of $a^{th}$ body ($a~=~1,~2$), $\omega_a~=~2\pi/P_{spin}$ is its angular velocity of rotation (here $a+1$ means modulo 2, $i.e.$ 2+1=1) and $n~=~2\pi/P_{orb}$. $x_1~=~m_1/M,~ x_2~=~m_2/M$ and $M~=~m_1+m_2$, $\bf k$ is the orbital angular momentum vector, $ \bf s_a$ is the spin vector and $\bf K_0$ is the line of sight vector.  Using eqn \ref{eq:per_adv}, the moment of inertia can be estimated from the observed value of $\dot{\omega}$ provided we know $P_{spin}$, $e$, $P_{orb}$, $m_1$ and $m_2$ (which can be estimated from timing analysis of radio data) of the binary components which is the case for J0737-3039 A and B.

On the other hand, the moment of inertia of a star of known mass can be theoretically calculated using a particular EoS. Bejger $et~al.$ \cite{bbh} calculated the moment of inertia for both PSR J0737$-$3039 A and B using 25 EoS of dense matter including an EoS for strange quark matter. Bejger $et~al.$ \cite{bbh} found $I_A$ to lie between $1.894 \times 10^{45} {\rm g~cm^2}$ and $0.726  \times 10^{45} {\rm g~ cm^2}$. They also checked that the moment of inertia estimated using Hartle's approximations (as described in Kalogera \& Psaltis \cite{kp}) differs from the exact calculation with rotating axisymmetric stellar model by not more than 1\%.  We re-performed the analysis and found $I_A$ to lie between $0.67-0.74  \times 10^{45}{\rm g~ cm^2}$. It is worthwhile to mention that the stiffest EoS (MFT17) chosen by Bejger $et~al.$  strongly overestimate  and the softest EoS (BPAL12) strongly underestimate nuclear matter incompressibility. So from the view point of a theoretician, the value of $I_A$ is likely to be somewhere in the middle region of figure 1 of Bejger $et~al.$ \cite{bbh}.

But the point to note here is that, in the expression of $\dot{\omega}$, the moments of inertia of both members of a DNS system are present (eqn \ref{eq:per_adv}) which means that we have one observable and two unknowns which is impossible to solve. But we have checked that the $\beta_{sa}$ term is significant only for  $P_{spin}<100~ ms$, so the term for J0737$-$3039B can be neglected and only the contribution from J0737$-$3039A remains. So determining the moment of inertia of PSR J0737$-$39B is impossible where measuring the same for PSR J0737$-$39A is possible by neglecting the spin-orbit coupling term due to PSR J0737$-$39B. This implies that this method of understanding the form of the matter building the star from pulsar timing analysis can be used for PSR J0737$-$39A, but not for PSR J0737$-$39B.  So we can not test the prediction of $I_B$ as claimed by Bejger $et~al.$ \cite{bbh}.

Even for PSR A, to measure the moment of inertia within $10\%$ of accuracy, we need to improve the accuracy in the knowledge of binary parameters to a significant extent. This possibility was studied recently by Iorio \cite{iorio} who concludes that  ``the possibility of reaching in a near future the required accuracy to effectively constrain $I_A$ to $10\%$ level should be, perhaps, considered with a certain skepticism".

\subsection{genesis of PSR J0737$-$39B}

One can consider an estimate of minimum progenitor
mass for PSR J0737$-$39B, keeping in mind the recent comment by
Stairs $et~al.$ \cite{stdkm}, that an important point usually
neglected is the contribution from gravitational binding energy
which is defined as \be E_{grav}~=~M_P-M_G \label{eq:grav} \ee
where $M_G$ is the gravitational mass and $M_P$ is the proper
mass. The increase in baryonic energy is given by \be
E_{bar}~=~\int^R_0 \frac{\Delta E(r) n(r)
dr}{\left[1-\frac{2GM(r)}{r}\right]^{1/2}} \label{eq:bar} \ee
where $\Delta E$ is the increase in energy per baryon due to the
conversion from ONeMg matter to strange quark matter, $n$ is the
number density, G is the gravitational constant, M and R are the
mass and radius of the star. So the mass of the progenitor will
be 
\be M_{progenitor}~=~M_{pulsar}+E_{tot}+\Delta m
\label{eq:mass}\ee 
where \be E_{tot}~=~E_{grav}+E_{bar}\ee 
and $\Delta m$ is the mass loss during the conversion process. PSR
J0737$-$3039B, which has $M_G~=~1.249 M_{\odot}$, corresponds to $M_P~=1.5187$ in our SS model \cite{brdd} and we found $E_{grav}~=~0.2697 M_{\odot}$ and $E_{bar}~=~ 0.0708M_{\odot} $
giving $E_{tot}~=~0.3405 M_{\odot}$. So $M_{progenitor} = 1.249+0.3405 + \Delta m=1.5895+ \Delta m$. This value is close to that estimated by Piran and Shaviv \cite{ps05} $i.e.~1.55 \pm 0.2~M_\odot$ with a low kick velocity 30 km/s. On the other hand, if J0737$-$3039 is a neutron star, we get using APR EoS \cite{apr}, $M_P=1.4357$ giving $E_{grav}=0.1867$ and $E_{bar}=0$ giving $M_{progenitor} = 1.249+0.1867 + \Delta m =1.4357 + \Delta m$ which is also not far from Piran and Shaviv's. Remember, detailed simulation is needed to know the exact value of $\Delta m$ which we hope will be done in the future. So it is very difficult to identify whether $J0737-3039$ B is a strange star or a neutron star from these arguments.

Another interesting outcome is that, in our model \cite{brdd}, a strange star with $M_G~=~1.249~ M_\odot$ has $1.9058 \times 10^{57}$ baryons while a neutron star with $M_G~=~1.249~ M_\odot$ has $1.6390  \times 10^{57}$ while a normal NeOMg star of mass $1.59~ M_\odot$ has $1.9060 \times 10^{57}$ number of baryons. We have assumed, of course, that the gravitational mass is the same as its baryonic mass for a normal star - as the general relativistic effects are very small. This estimate shows that if $J0737-3039$ B is a strange star then the baryon loss is negligible ($0.01\%$) implying a small $\Delta m$. On the other hand, if $J0737-3039$ B is a neutron star then the baryon loss is larger ($14\%$) implying larger values for $\Delta m$. Small baryon loss (as obtained in the first case) supports Piran and Shaviv's model with small kick.  So this study would suggest PSR B to be a strange star.

\section{Conclusions and summary}

We have discussed how the double pulsar system can help us to constrain the dense matter EsoS.  Radio observations (timing analysis) may help us to constrain the EoS of PSR A in future and may allow us to deduce whether it is a strange star or a neutron star.

For PSR B, as this is not possible in the present situation, we can use indirect methods, $e.g.$ modeling its formation scenario from the present day observed parameters. 
The small mass of the star might be an indication that it is a strange star, since in the formation of such a star, in addition to the gravitational binding energy loss, mass is also lost in the form of baryonic energy. Although this is suggestive, more modeling is required to support such a conclusion.

\section{Acknowledgments}

The authors thank Ignazio Bombaci and Alak Ray for exciting
discussion, Bart Willems for helpful comments and Tomasz Bulik for
sending their moment of inertia results. MD thanks DST, Govt. of India, for Ramanna Fellowship. JD also thanks DST for the Ramanna project for sustenance.

\end{document}